\documentclass[a4paper,fleqn,usenatbib]{mnras}

\pdfoutput=1
\usepackage{newtxtext,newtxmath}
\usepackage{multirow}

\usepackage[T1]{fontenc}
\usepackage{ae,aecompl}

\usepackage{graphicx}	
\usepackage{amsmath}	
\usepackage{amssymb}	

\title[Swift data hint at a binary SMBH candidate]{Swift data hint at a binary Super Massive Black Hole candidate at sub--parsec separation}

\author[Severgnini et al.]{P. Severgnini,$^{1}$\thanks{E-mail: paola.severgnini@brera.inaf.it}
C. Cicone, $^{1}$
R. Della Ceca, $^{1}$
 V. Braito,$^{1,2}$
A. Caccianiga, $^{1}$
L. Ballo,$^{3}$
\newauthor
S. Campana, $^{1}$
A. Moretti, $^{1}$
V. La Parola,$^{4}$
C. Vignali,$^{5,6}$
 A. Zaino,$^{7}$
G. A. Matzeu, $^{1,8}$
\newauthor
 M. Landoni $^{1}$
\\
\\
$^{1}$ INAF -- Osservatorio Astronomico di Brera, via Brera 28, I-20121, Milano, Italy \& via Bianchi 46, I-23807, Merate, Italy\\
$^{2}$ Department of Physics, University of Maryland, Baltimore County, Baltimore, MD 21250, USA\\
$^{3}$ XMM-Newton Science Operations Centre, ESAC/ESA, PO Box 78, E-28691 Villanueva de la Ca\~{n}ada, Madrid, Spain \\
$^{4}$ INAF -- Istituto di Astrofisica Spaziale e Fisica Cosmica di Palermo, via Ugo La Malfa, I-90146, Palermo, Italy\\
$^{5}$ Dipartimento di Fisica e Astronomia, Alma Mater Studiorum, Universit\`a degli Studi di Bologna, Via Gobetti 93/2, I-40129 Bologna, Italy\\
$^{6}$ INAF -- Osservatorio Astronomico di Bologna, Via Gobetti 93/3, I-40129 Bologna, Italy\\
$^{7}$ Dipartimento di Matematica e Fisica, Universit$\grave{a}$ degli Studi Roma Tre, via della Vasca Navale 84, I-00146, Roma, Italy\\
$^{8}$ European Space Agency (ESA), European Space Astronomy Centre (ESAC), E-28691, Villanueva de la Ca\~{n}ada, Madrid, Spain\\
}

\date{Accepted 2018 June 25. Received 2018 June 25; in original form 2017 December 14}

\pubyear{2015}

\begin{document}
\label{firstpage}
\pagerange{\pageref{firstpage}--\pageref{lastpage}}
\maketitle

\begin{abstract}
Dual/binary Supermassive Black Hole (SMBH) systems are the inevitable consequence of the current $\Lambda$ Cold Dark Matter
cosmological paradigm. In this context, we discuss here the properties of MCG+11-11-032, a local ({\it z}=0.0362)
Seyfert 2 galaxy. 
This source was proposed as a dual AGN candidate 
on the basis of the presence of  double-peaked [OIII] emission lines in its optical spectrum. MCG+11-11-032
 is also an X-ray variable source and was observed several times by the {\it Swift} X-ray Telescope (XRT) on time scales from days to years.
 In this work, we analyze the SDSS-DR13 spectrum and find evidence for 
 double-peaked profiles  in  all the strongest narrow emission  lines.
 We also study the XRT light curve and unveil the presence of an alternating behavior of the intrinsic 0.3-10 keV flux, while the 
 123-month  {\it Swift} BAT light curve supports the presence of 
almost regular  peaks and dips almost every 25 months. In addition, the XRT spectrum suggests for the presence of two  
narrow emission lines  with rest-frame energies of {\it E}$\sim$6.16 keV and  
{\it E}$\sim$6.56 keV.
 Although by considering only the optical emission lines, different physical mechanisms may be invoked to explain the kinematical properties,
 the X-ray results  are most naturally explained by the presence of a binary SMBH in the center of this source. In particular, we evidence a remarkable agreement between 
 the  putative SMBH pair orbital velocity derived from the BAT light curve and  the velocity offset derived by  the rest--frame $\Delta E$ between the two X-ray line peaks
in the XRT spectrum (i.e. $\Delta v\sim$0.06{\it c}).  
\end{abstract}




\section{Introduction}
The search for and the characterization of the dual (kpc scale
separation) and binary (pc separation) active supermassive black hole (SMBH, {\it M}$_{\mathrm{BH}} >$10$^6$ {\it M}$_{\odot}$)
population is a hot topic of current astrophysics, given its relevance to
understand galaxy formation and evolution. 
Since it is now clear that the most
massive galaxies should harbor a central SMBH \citep{kor95, fer05}, the formation  of 
dual/binary SMBH systems is the inevitable consequence of the current $\Lambda$CDM
cosmological paradigm, in which galaxies grow hierarchically
through minor and major mergers.
The dynamical evolution of dual/binary SMBH systems within the merged galaxy, and
their interaction with the host (via dynamical encounters and feedback
during baryonic accretion onto one or both SMBHs) encode crucial information
about the assembly of galaxy bulges and SMBHs. 
In addition to this, if the binary SMBHs eventually
coalesce, they will emit gravitational waves that  could be detected with incoming low frequency
gravitational wave experiments \citep{eno04}.\\
Although dual/binary AGN are a natural outcome of  galaxy mergers, the
number of confirmed dual/binary AGN is still too low when compared with
  model expectations  \citep[e.g.,][]{spr05,hop05}. Indeed, directly observing SMBHs during different merger stages is still a challenging task not only because
of  the stringent resolution requirements, but also because of  the intrinsic  difficulty
in identifying SMBHs. In  late merger stages, SMBHs are expected to be
embedded in a large amount of dust and gas and thus strongly obscured and
elusive both in the UV and optical bands. Only a few tens of dual
SMBHs at $<$10 kpc separation have been confirmed \cite[see][and
references therein]{mcg15}  and only a
few definitive sub-kpc binary SMBHs have been discovered and studied so far 
\citep[e.g.,][] {rod06,val08,bor09}. 

The search for double-peaked  optical emission lines emerging from two separate 
narrow-line regions (NLRs)  of two SMBHs has been proposed
 as a method to select dual AGN candidates on kpc/sub-kpc scales  \citep[e.g.,][]{wan09}.
The catalogs of  \cite{wan09}, \cite{liu10}, and \cite{smi10}  provide about three hundreds of unique candidate dual AGN, selected from the Sloan Digital Sky Survey
Data Release 7 \citep[SDSS--DR7,][]{aba09}
 as spectroscopic
AGN with a double-peaked [OIII]$\lambda$5008\AA{} line (redshift range 
between 0.008$<${\it z}$<$0.686).
However, the 3" diameter of the SDSS fiber does not discern whether the double-peaked optical emission lines are due
to dual kpc-scale nuclei or  to kinematical  effects occurring within a single AGN, e.g.  jet-cloud interactions 
\citep{hec84,gab17}, a rotating, disk-like NLR \citep{xu09}, or the combination of a blobby NLR and extinction effects \citep{cre10}. 

In this paper we present  and discuss the optical  and X-ray  properties of MCG+11-11-032, a radio-quiet optical Seyfert 2 galaxy at {\it z}=0.0362. 
Besides being a  dual AGN candidate on the basis of its SDSS--DR7 optical spectrum
 characterized by  double-peaked [OIII] emission lines \citep{wan09}, MCG+11-11-032 is also 
an X-ray variable source \citep{bal15a}.
It belongs to the  all-sky survey {\it Swift}-BAT catalogues \citep{bau13,cus14} and 
was observed several times by the  {\it Swift} X-ray Telescope \citep[XRT;][]{bur05}  on
time scales from years to days.
In Section 2, after summarizing previous results,
we present our new analysis of the SDSS--DR13 spectrum of MCG+11-11-032. The analysis of the  XRT light-curve  and spectra,
along with the 123-month BAT light curve, are presented in Section 3.
In Section 4, we combine the optical spectroscopic information with the X-ray results to 
 discuss the most plausible  physical scenarios acting in MCG+11-11-032. Section 5 presents our conclusions.

Throughout the paper we assume a flat $\Lambda$CDM cosmology with H$_0$=69.6 km s$^{-1}$ Mpc$^{-1}$, $\Omega_{\Lambda}$=0.7 and 
$\Omega_{\rm M}$=0.3. Errors are given at 68 per cent confidence level unless otherwise specified, i.e. Sect. 3.

\section{SDSS spectrum}

MCG+11-11-032 (SDSS J085512.54+642345.6) 
belongs to a sample of 87 SDSS--DR7 type 2 AGN with double-peaked [OIII] profiles
selected and analyzed by \cite{wan09}.
The obscuration of the active nucleus allowed the authors  to determine the redshift
of the host galaxy  ({\it z}=0.03625$\pm$1$\times$10$^{-5}$) through the stellar absorption lines and investigate the 
properties of the nebular emission lines.
They fitted the [OIII] line with a blue-shifted  ($\Delta \lambda_{\rm b}$=-2.52$\pm$0.07 \AA{}) and a red--shifted ($\Delta \lambda_{\rm r}$=2.26$\pm$0.06 \AA{}) component, neither of which is at the systemic velocity of the host galaxy.
The corresponding fluxes and luminosities  of the blue-shifted and red-shifted components by \cite{wan09} are:
 $F_{\rm[OIII] blue}$=[338$\pm$11]$\times$10$^{-17}$ erg cm$^{-2}$ s$^{-1}$, $L_{\rm[OIII] blue}$=1.04$\times$10$^{40}$ erg  s$^{-1}$    and 
$F_{\rm[OIII] red}$=[372$\pm$11]$\times$10$^{-17}$ erg cm$^{-2}$ s$^{-1}$, $L_{\rm[OIII] red}$=1.14$\times$10$^{40}$ erg  s$^{-1}$.

 \cite{com12}  performed follow--up long-slit observations with the Blue Channel Spectrograph
on the MMT 6.5 m telescope. They observed
the object at two different position angles (one along the isophotal position angle of the major axis
of the host galaxy and the other one along the orthogonal axis) in order to determine the full spatial separation of the two emission line components.
They fit two Gaussian components to the continuum subtracted [OIII]$\lambda$5007\AA{} emission line  and
 found a velocity offset between the  [OIII]$\lambda$5007\AA{} peaks of  $v$=275$\pm$4 km s$^{-1}$.
They measured the  angular  and  physical projected spatial  
offset between the two [OIII]$\lambda$5007\AA{} emission features and found 0.77$\pm$0.04 arcsec and
0.55$\pm$0.03 {\it h}$^{-1}_{70}$ kpc, respectively, with a PA=61.0$^\circ$$\pm$2.2. 
By checking the long-slit 
spectra {\it by eye},  \cite{com12}  classified  the AGN emission components as 
spatially "extended", meaning that not all the emission line components appear spatially compact at  the 
position angles observed. No additional information has been provided by the authors about the emission line profiles at each position angle. 

\begin{figure}
        \includegraphics[scale=0.38,clip=true,trim=0cm 5cm 0cm 3cm]{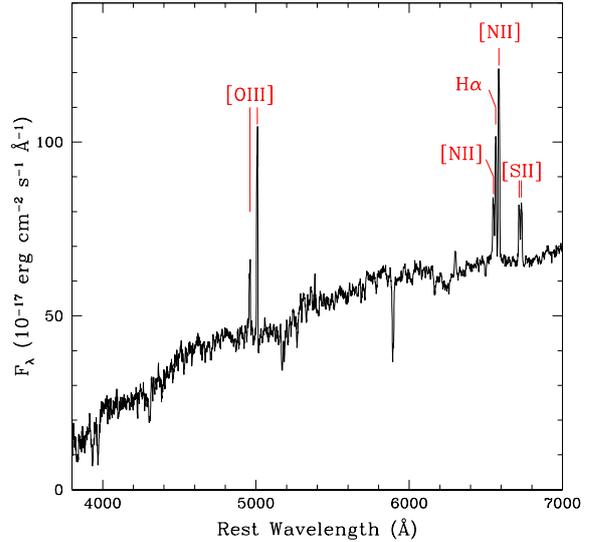}
    \caption{Optical (SDSS--DR13) spectrum in the rest frame of the MCG+11-11-032 host galaxy
    ({\it z}=0.036252). The strongest emission lines are labelled.}
    \label{sdss_spectrum}
\end{figure}

\begin{figure}
\centering
\vskip -0.2truecm
        \includegraphics[scale=0.3,clip=true,trim=0cm 5cm 0cm 3cm]{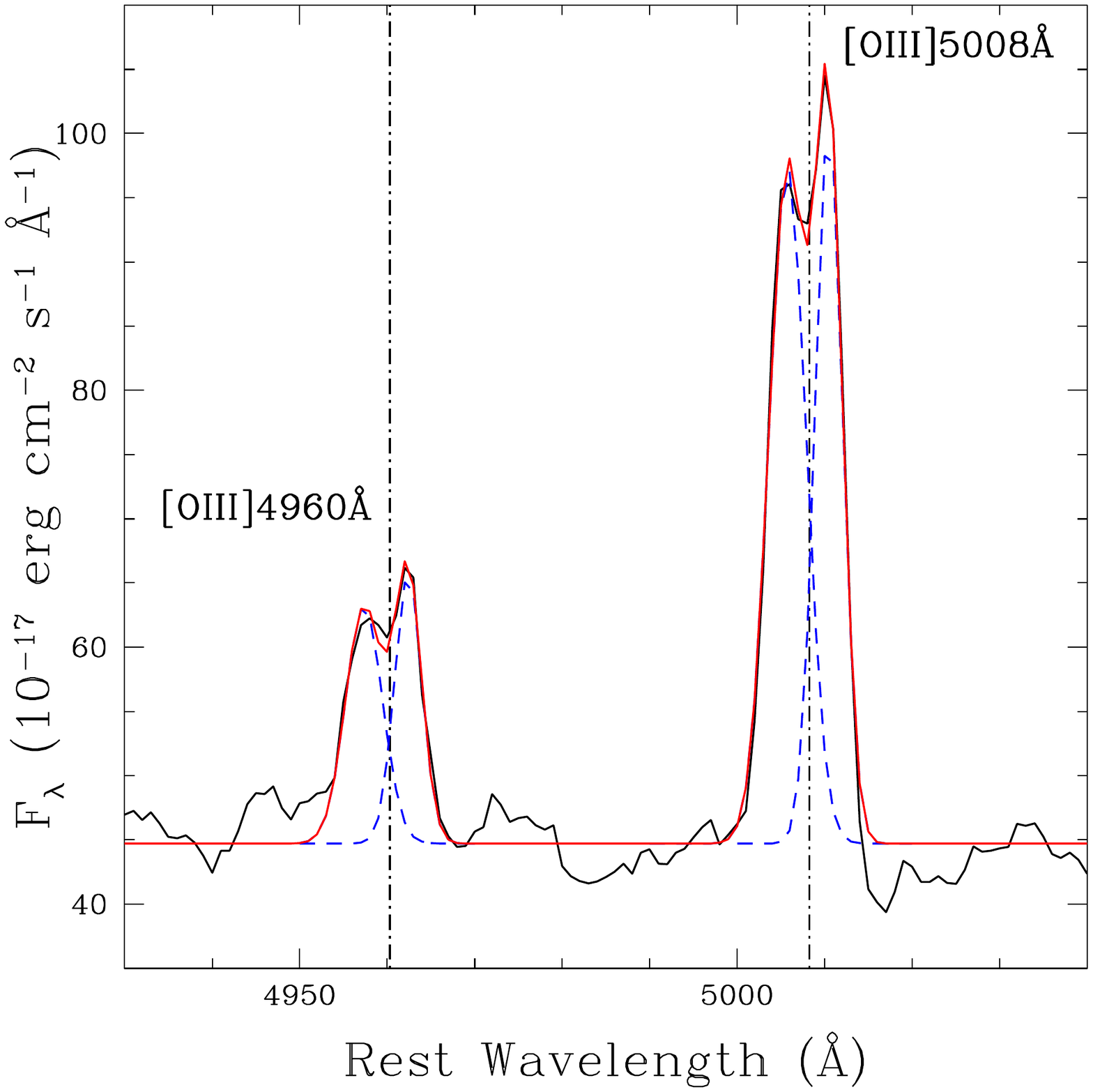}
        \includegraphics[scale=0.3,clip=true,trim=0cm 5cm 0cm 3cm]{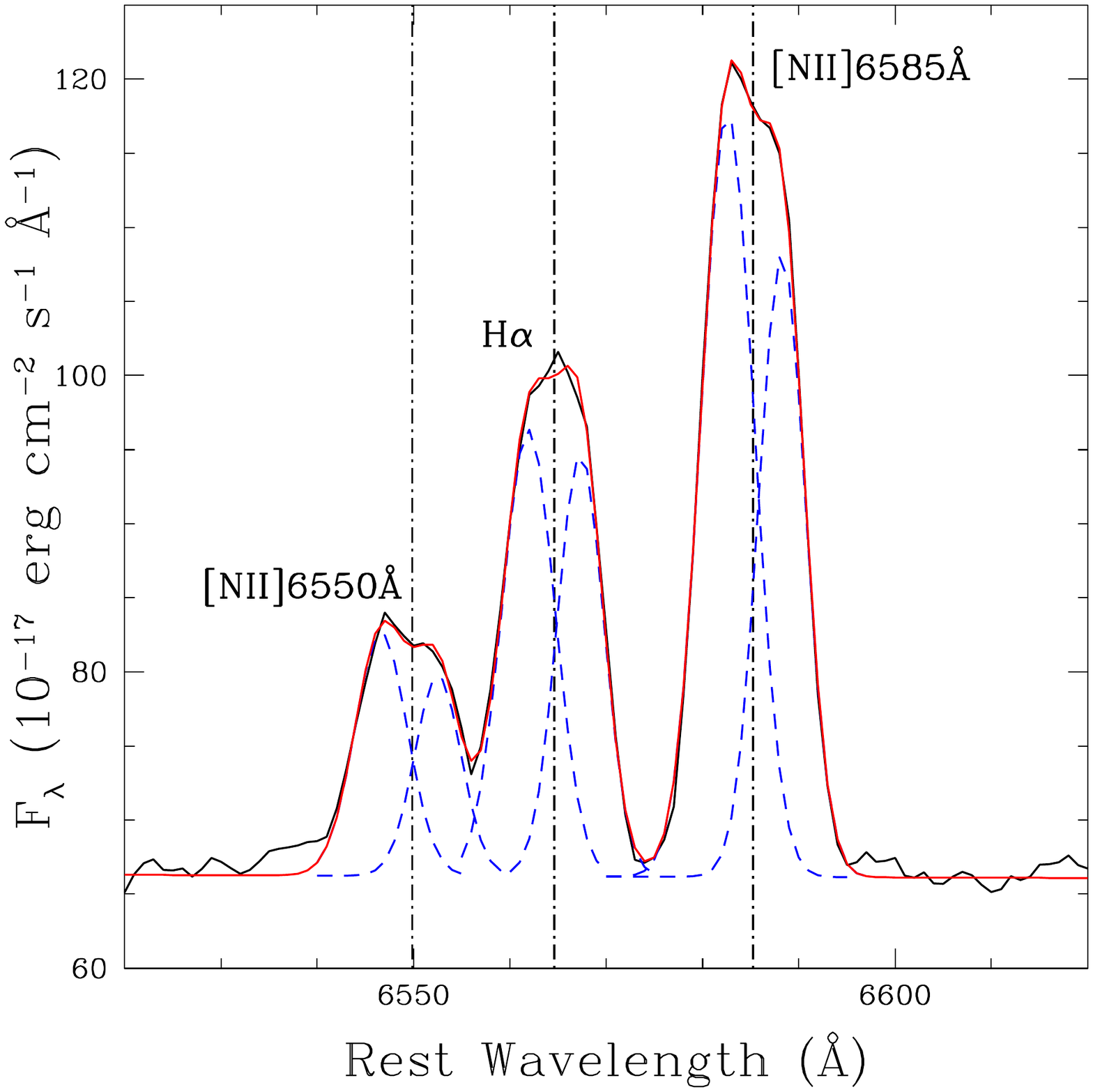}
        \includegraphics[scale=0.3,clip=true,trim=0cm 5cm 0cm 3cm]{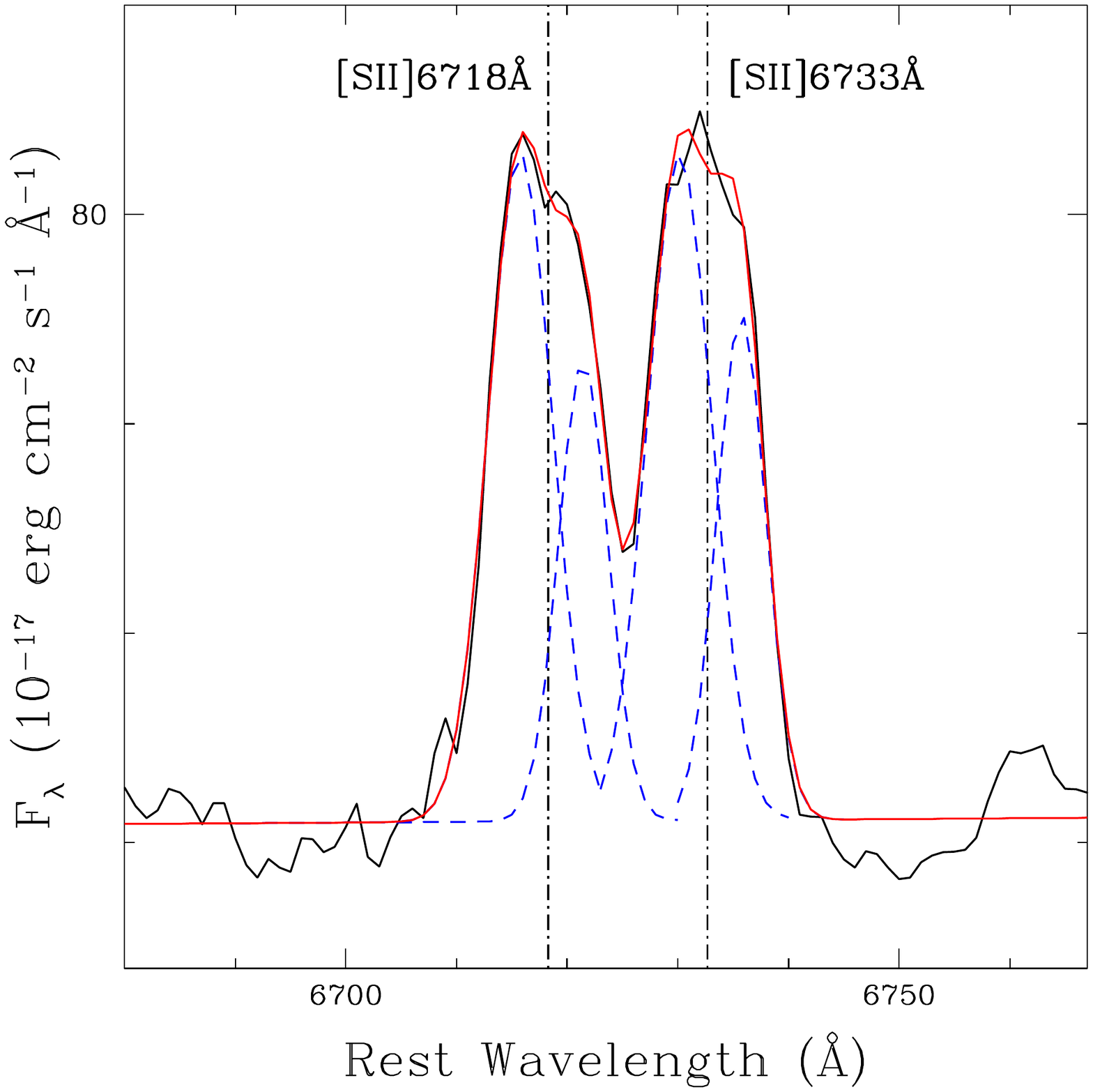}
    \caption{MCG+11-11-032 rest-frame spectrum around the [OIII] (top panel),  H$\alpha$+[NII] (middle panel) and
  [SII] (bottom panel)  regions. Wavelengths are in vacuum. Red solid lines represent the total fit, while blue dashed lines are the gaussian components. Black dot--dashed vertical lines mark the position of the
  line at the systemic velocity of the host galaxy.}
    \label{sdss_lines}
\end{figure}

\begin{table*}
 \begin{minipage}[t]{1\textwidth}
        \begin{center}
  \caption{Properties of the strongest emission line components in the SDSS--DR13 spectrum of MCG+11-11-032.}
  \label{sdss_results}
  \begin{tabular}{cccccccc}
   \hline
 Spectral line & $\lambda_{\rm peak}$ & {\it FWHM} & $\Delta \lambda_{\rm peak}$ &$\Delta {v}_{\rm peak}$   &  $v_{\rm red} - v_{\rm blue}$& Flux & Luminosity     \\
                     &   [\AA{}]                       &  [km/s]       &    [\AA{}]                                 &       [km/s]                       &       [km/s]                       &        [10$^{-17}$ erg/cm$^{2}$s ]  &   [10$^{39}$ erg/s]\\
         (1)       &      (2)   &   (3)     &     (4)    &   (5)      &   (6)                                      &    (7) & (8)   \\            
                     \hline
                     \hline
 \multirow{ 3}{*}{$\rm {[OIII]}{\lambda 4960.295}$\AA{}} & 4957.40$\pm$0.23 &    300$^a$  &   -2.93$\pm$0.33 & -177$\pm$20  &\multirow{ 3}{*}{305$\pm$17} & 140$\pm$11 & 4.3$\pm$0.3\\
 ~~~\\
                                                                                         & 4962.41$\pm$0.17 & 222$^b$  & 2.11$\pm$0.24 & 128$\pm$15  & & 118$\pm$11 & 3.6$\pm$0.3\\
 ~~~\\
 ~~~\\
\multirow{ 3}{*}{ $\rm {[OIII]}{\lambda 5008.239}$\AA{}}  & 5005.75$\pm$0.15 & 300$\pm$17 & -2.49$\pm$0.21 & -149$\pm$13 & \multirow{ 3}{*}{283$\pm$11} & 402$\pm$24 & 12.3$\pm$0.7\\
 ~~~\\
                                                                                & 5010.48$\pm$0.11 & 222$\pm$11 & 2.24$\pm$0.16   &  134$\pm$9 &  & 317$\pm$39 &9.7$\pm$0.6\\
\hline
\hline
 ~~~\\
\multirow{ 3}{*}{  $\rm {[NII]}{\lambda 6549.86}$\AA{}}  & 6546.61$\pm$0.19 & 294$^c$  & -3.25$\pm$0.27 & -149$\pm$12 & \multirow{ 3}{*}{271$\pm$13} & 142$\pm$8 & 4.4$\pm$0.3 \\                                                                                                                                                                                                                                                                                                                                             
~~~\\
                                                                               & 6552.53$\pm$0.20  & 260$^d$   & 2.66$\pm$0.28   &  122$\pm$13 &  & 104$\pm$23 &  3.2$\pm$0.7\\
 ~~~\\
 ~~~\\
\multirow{ 3}{*}{ $\rm {H\alpha}{\lambda 6564.614}$\AA{}}  & 6561.89$\pm$0.17 & 294$^c$  & -2.70$\pm$0.41 & -124$\pm$18 &\multirow{ 3}{*}{248$\pm$11}  &  262$\pm$32 & 8.0$\pm$0.8\\
 ~~~\\
                                                                                    & 6567.32$\pm$0.16  & 260$^d$   & 2.71$\pm$0.26   &  124$\pm$12 & & 219$\pm$68 & 6.7$\pm$2.3\\
~~~\\
 ~~~\\
\multirow{ 3}{*}{ $\rm  {[NII]}{\lambda 6585.27}$\AA{}}  &6582.59$\pm$0.15 & 294$\pm$10 & -2.68$\pm$0.21 & -122$\pm$13 & \multirow{ 3}{*}{259$\pm$10} & 449$\pm$38 & 13.8$\pm$1.2\\
~~~\\
                                                                               &6588.26$\pm$0.15  & 260$\pm$9   &2.99$\pm$0.25  &  136$\pm$11  & & 324$\pm$48 & 9.9$\pm$01.5 \\
\hline
\hline
~~~\\
\multirow{ 3}{*}{ $\rm  {[SII]}{\lambda 6718.29}$\AA{}}  & 6715.77$\pm$0.33 & 303$\pm$24   & -2.52$\pm$0.47 & -112$\pm$21 & \multirow{ 3}{*}{254$\pm$20}& 147$\pm$15 & 4.5$\pm$0.4\\
~~~\\
                                                                              & 6721.45$\pm$0.30  &  233$\pm$21   &3.16$\pm$0.42   &  141$\pm$19 &  & 77$\pm$12 & 2.4$\pm$0.4\\
~~~\\
~~~\\
\multirow{ 3}{*}{ $\rm  {[SII]}{\lambda 6732.68}$\AA{}} & 6730.15$\pm$0.32 &  303$^e$  & -2.53$\pm$0.45 & -113$\pm$20 & \multirow{ 3}{*}{250$\pm$21} & 147$\pm$9  & 4.5$\pm$0.3\\
 ~~~\\
                                                                             & 6735.76$\pm$0.32  &   233$^f$  &3.08$\pm$0.45   &  137$\pm$20 & &  85$\pm$12  & 2.6$\pm$0.4 \\
\hline
\end{tabular}
\end{center}

{\bf Notes.}  Col. (1): Spectral lines (rest--frame vacuum wavelengths). 
Col. (2): Rest-frame wavelengths of the peaks of the blue and red-shifted emission line components.
Col. (3): $FWHM$, not corrected for the instrumental resolution, of the blue and red-shifted emission line components. 
Col (4): Doppler shifts of the blue and red emission line components. 
Col. (5): Line of sight velocity offsets of the blue and red-shifted components. 
Col. (6): Line of sight velocity offset between red and blue peaks. 
Col. (7)-(8): Fluxes and luminosities, corrected for Galactic extinction, of the blue and red-shifted components. \\
$^a$ fixed to be equal to $FWHM$ of  the $\rm {[OIII]}{\lambda 5008.239}$\AA{}  blue component. $^b$ fixed to be equal to $FWHM$ of  the $\rm {[OIII]}{\lambda 5008.239}$\AA{}  red component. $^c$ fixed to be equal to $FWHM$ of  the $\rm {[NII]}{\lambda6585.27}$\AA{} blue component. $^d$ fixed to be equal to $FWHM$ of  the $\rm {[NII]}{\lambda6585.27}$\AA{} red component. $^e$ fixed to be equal to $FWHM$ of  the $\rm  {[SII]}{\lambda 6718.29}$\AA{}  blue component. $^f$fixed to be equal to $FWHM$ of  the $\rm  {[SII]}{\lambda 6718.29}$\AA{}  red component.
  
\end{minipage}
\end{table*}

To further investigate the presence of double peaked emission lines, 
we analyzed the SDSS--DR13 spectrum.
We examined the
properties of all the prominent emission lines detected in the spectrum (see Fig.~\ref{sdss_spectrum}), instead of
considering  only the  [OIII]$\lambda$5007\AA{} (as it was done by previous authors).
To this end, we analyzed the spectral region around  the [OIII],  H$\alpha$+[NII], and 
[SII] lines by fitting  the rest-frame 4910-5070 \AA{}, 6500-6660 \AA{} and 6660-6715 \AA{}
ranges, avoiding the regions where absorption lines are present.
We used a combination of a power-law continuum and two Gaussian components for each transition.
In this procedure, for each Gaussian component there are three independent parameters that need to be determined: 
the position of the line peak ($\lambda_{\rm peak}$), the line broadening ($\sigma$) and
the normalization of the line ($N$).
The central line positions and their relative intensities were set to be all independent.
We note, however, that the emission line ratios derived by our fit are in good agreement with the expected theoretical values, e.g.
3:1 for the two [OIII] lines.
Since the narrow lines observed in this source should have the same  physical origin, i.e. they arise from the NLRs, 
we performed a first fit in which the widths of all the Gaussian components were forced to have the
same value in units of km s$^{-1}$.
However, since for  forbidden transitions larger  
critical electron densities for de-excitation (i.e. {\it N$_{\rm ec}$}[OIII]$>${\it N$_{\rm ec}$}[NII]$>${\it N$_{\rm ec}$}[SII], see \citeauthor{ost91}  \citeyear{ost91}) can
broaden the line profile, we left the line widths to be independent between the three different spectral 
ranges considered here. 
Each emission line resulted to be well fitted by two narrow Gaussian components.
In particular, for the two [OIII]$\lambda$5007\AA{} lines our results
are in good agreement with \cite{wan09} and  \cite{com12}
in terms of peak offsets and line fluxes. 
 As a second and final  step, we re-fitted the data by allowing
the widths of the bluer Gaussian components to be 
 independent  with respect to the red ones.
The results of this second fit are shown in Fig.~\ref{sdss_lines} and reported in Table~\ref{sdss_results} together with 1$\sigma$ statistical errors.
The addition of  a third  Gaussian  
to account for a possible component at the systemic velocity is not requested by the fit.
We note that outflows/jets in integrated galaxy spectra, where the individual outflowing clouds are not
resolved by the observations,  tend to appear in the form of fainter -- and broader -- possibly asymmetric emission line components
superimposed on the narrow emission line centered at the systemic velocity  \citep[e.g.][]{har14,har16}. 
The lack of such typical outflow components 
does not support the presence of strong ionized outflows or jets detectable in the SDSS13 spectrum.
The absence of a strong radio jet is also in agreement with the non detection of  MCG+11-11-032 
down to about 0.5 mJy (3$\times$rms) at 1.4 Ghz in the VLA FIRST survey \citep{bec95}. 
However, we note that, although  strong radio jets can be ruled out, the presence of faint jets cannot be discarded with the available optical and radio data.

All the results reported here were obtained by combining a power-law continuum plus narrow Gaussian components, and we have verified that they are not dependent (within the 1$\sigma$ uncertainties) on the stellar continuum subtraction.

In summary, the value reported in Table~\ref{sdss_results} are in agreement with 
 the results obtained by \cite{wan09} and \cite{com12} for the [OIII] line. 
 In addition, we confirm  the  presence of
 double-peaked profiles in all the  optical nebular emission lines detected in the MCG+11-11-032  SDSS--DR13 spectrum. The widths of all the 
 blue and red components are consistent, within 3$\sigma$ uncertainties, with each other across 
 the full spectral range, though the width of the red components is systematically smaller.
 We found a significant (more than 3$\sigma$) offsets between the blue and red peaks with respect to the systemic redshift. This implies that none of these components is consistent with the nominal systemic velocity of the host galaxy (derived from the stellar absorption
lines).  Finally,  the offset ratios ($\Delta \lambda_{\rm blue}$/$\Delta \lambda_{\rm red}$) of the two components of H$\alpha$, [NII], and
[SII] are consistent with that of [OIII], and they are of the order of one, i.e.  the wavelength shifts and the corresponding velocity offsets 
between the blue and red peaks are similar for all the nebular emission lines.

Following statistical argument  originally proposed by \cite{wan09}, these results can be explained by
 the presence  of two distinct NLRs that, on the basis of their projected physical separation 
estimated by \cite{com12}  (0.55$\pm$0.03 {\it h}$^{-1}_{70}$ kpc), may be related to  two different AGN. The extended structures detected by  the same authors
could be produced by either gas kinematics in the two NLRs or faint outflows/jets associated to  one or both SMBHs.

We note, however, that the line properties reported in Table~\ref{sdss_results}  as well as the presence of spatially extended components can be also explained  by alternative physical scenarios. Firstly, as quoted before, the presence of a single SMBH associated with faint jet activity  can not be  excluded based on the available optical and radio data. Secondly, the velocity offsets reported in Table~\ref{sdss_results} are 
fully  consistent with rotation velocities measured in
nearby galaxies \citep[see e.g.][]{sof01}.
This implies that the double-peaked emission lines observed for  MCG+11-11-032 could be produced by gas kinematics related to a single NLR \citep[see e.g.][]{ble13,fu12}, partially or totally tracing the kinematics of the almost edge-on \citep[{\it b/a}=0.45 by][]{kos11} host galaxy disk at sub-kpc scale. Another intriguing possibility, corroborated by the X-ray data presented in the next Section, is that we are observing
gas kinematics effects produced by a single NLR  ionized by two SMBHs near to the coalescence phase, i.e.
a sub-pc scale SMBH pair (see Section 4).

\section{X-ray data}
\begin{table}
\scriptsize
 \begin{center}
  \caption{MCG+11-11-032: XRT  observation log. The observations are ordered
 on  the basis of the observation starting date.}
  \label{xobs_log}
  \begin{tabular}{@{}lccc@{}}
  \hline
  \hline
Obs. ID & Obs.  Start date& Net Counts   &  Net Exp. Time\\
   (1)      &     (2)                &   (3)               & (4) \\
 \hline
{\it Archival data} & & & \\
 \hline
00038045001  & 2008-11-20 & 54 & 4957\\
 00038045002  & 2009-01-28 & 135 & 5022 \\
 00090163002  & 2009-09-06 & 42  & 5594\\
 00084954001  & 2015-01-23 & 12  & 457\\
 00084954002  & 2015-02-02 &  16 & 622 \\
 00084954005  & 2015-03-24 &  9 & 1249\\
  \hline
{\it Our own monitoring} &&& \\
 \hline
 00034134001   & 2015-12-14 & 94   & 3236 \\
 00034134002   & 2015-12-15 & 106 & 4430\\
 00034134003   & 2015-12-15  & 126 & 7317 \\
 00034134004   & 2015-12-19  & 34 & 899 \\ 
 00034134005   & 2015-12-19  & 460 & 15890\\
 00034134006   & 2015-12-24 & 230  & 8346 \\
 00034134007   & 2015-12-25 & 352 & 11050\\
 00034134008   & 2015-12-29 & 187 & 8129 \\ 
 00034134009   & 2015-12-30 & 161 & 5866\\ 
 00034134010   & 2016-01-10 & 272 & 7337\\
 00034134011   & 2016-01-12  & 182 & 7280\\
 00034134012   & 2016-01-13 & 26 & 942\\
 00034134013   & 2016-01-14 & 51 & 2527\\
 00034134014   & 2016-01-14 & 265 & 16810\\
 00034134015   & 2016-01-19 & 65 & 3326\\
 00034134016   & 2016-01-20 & 225 & 8259\\
 00034134017   & 2016-01-21 & 23 & 1096\\
 00034134018   & 2016-01-22 & 21 & 1056\\
 00034134019   & 2016-01-24 & 107 & 6171\\
 00034134020   & 2016-01-27 & 17 & 942\\
 00034134021   & 2016-01-28 & 218 & 10340\\
 00034134022   & 2016-01-29 & 263 & 15400\\
 00034134024   & 2016-02-03 & 99 & 3906\\ 
 00034134025   & 2016-02-03 & 101 & 3591\\
 00034134026   & 2016-02-05 & 338 & 11530\\
  \hline
{\it Archival data} && & \\
 \hline
00080403001   & 2016-02-18 & 60  & 1826 \\
\hline
\end{tabular}
\end{center}
{\bf Notes.}  Col. (1): XRT observational ID. (2)  Start date of the observation. (3) [0.3-10 keV] net counts as derived by 
X-ray spectral analysis. (4) Nominal exposure time in unit of second.
\end{table}

We  monitored MCG+11-11-032  with the {\it Swift}-XRT telescope as part 
of a project aiming at studying the X-ray variability  on different time scales  \citep{bal15a}.
The observations, performed during the {\it Swift} Cycle-12 (P. I. Severgnini), started on  2015 December 14 and ended on 2016
February 5, covering $\sim$54 days, for a total  exposure time of $\sim$166 ks.
The data were taken using XRT in the standard PC-mode (Target ID=00034134). 
The observation log is reported in Table~\ref{xobs_log}; we note that the  observations scheduled
during the segment 00034134023 were not completed, explaining the absence of this latter in the observation log. 
A further archival observation (Target ID=00080403), was performed just after our own monitoring and it was  considered in our analysis.
Before our  daily monitoring, MCG+11-11-032 was observed different times by {\it Swift} on 
month/year time scales. 
The relevant information about the previous XRT  observations
in which the source was detected with a S/N$>$3 in the 0.3-10 keV range is  reported  in the
first part of Table~\ref{xobs_log} (Target ID=00038045, 00090163 and 00084954).

For each observation, we extracted the images, light curves, and spectra, including the background and ancillary response files, 
 using the on-line XRT data product generator\footnote{http://www.swift.ac.uk/user\_objects} \citep{eva07, eva09}. 
The effects of the damage to the CCD and automatic 
readout-mode switching were handled and the appropriate spectral response files were identified in the calibration database.
The source appears to be point-like in the
XRT image and we do not find any significant evidence of pile-up.

\begin{figure*}
        \includegraphics[scale=0.4,clip=true,trim=0cm 5cm 0cm 3cm]{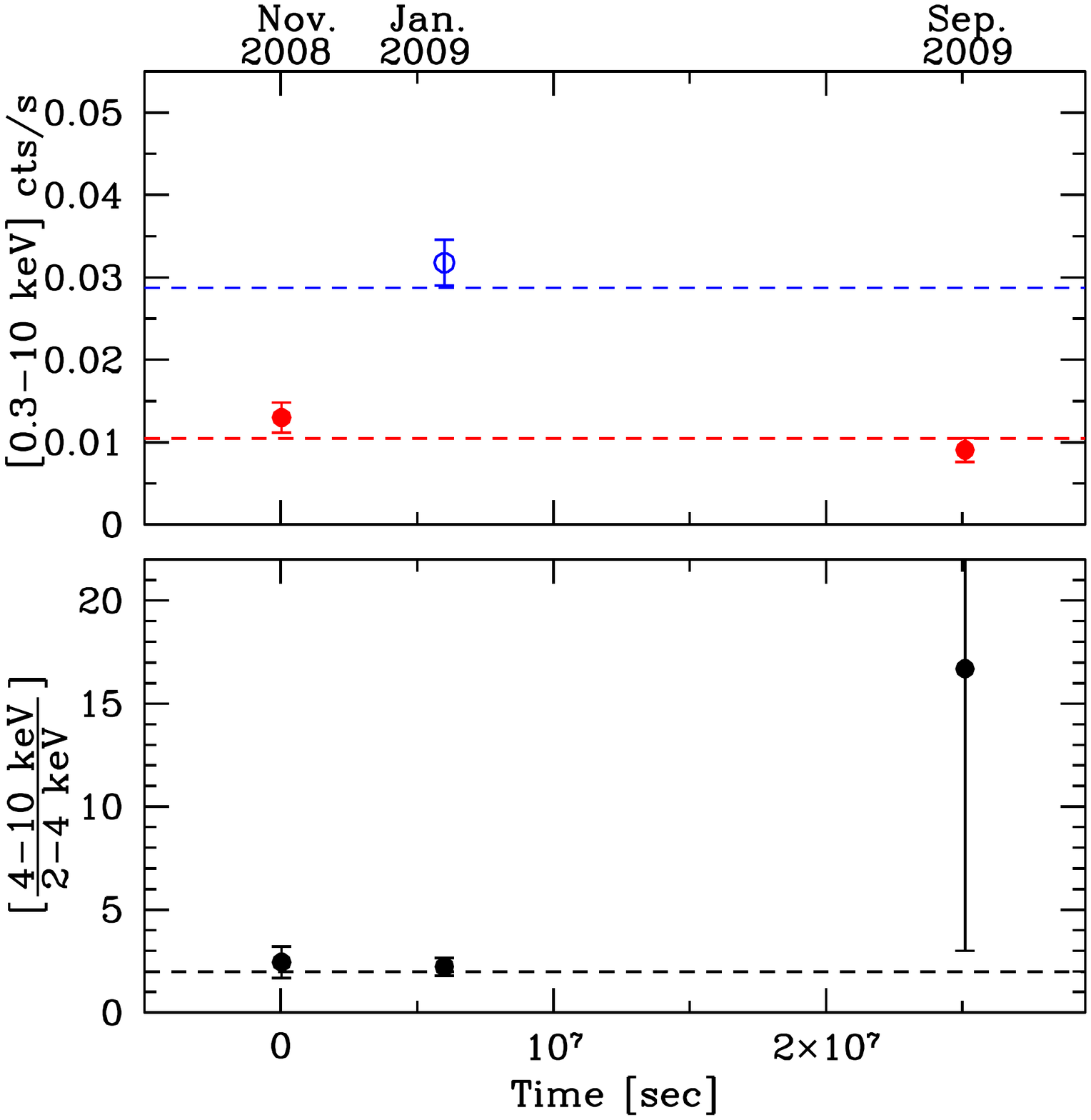}
        \includegraphics[scale=0.4,clip=true,trim=0cm 5cm 0cm 3cm]{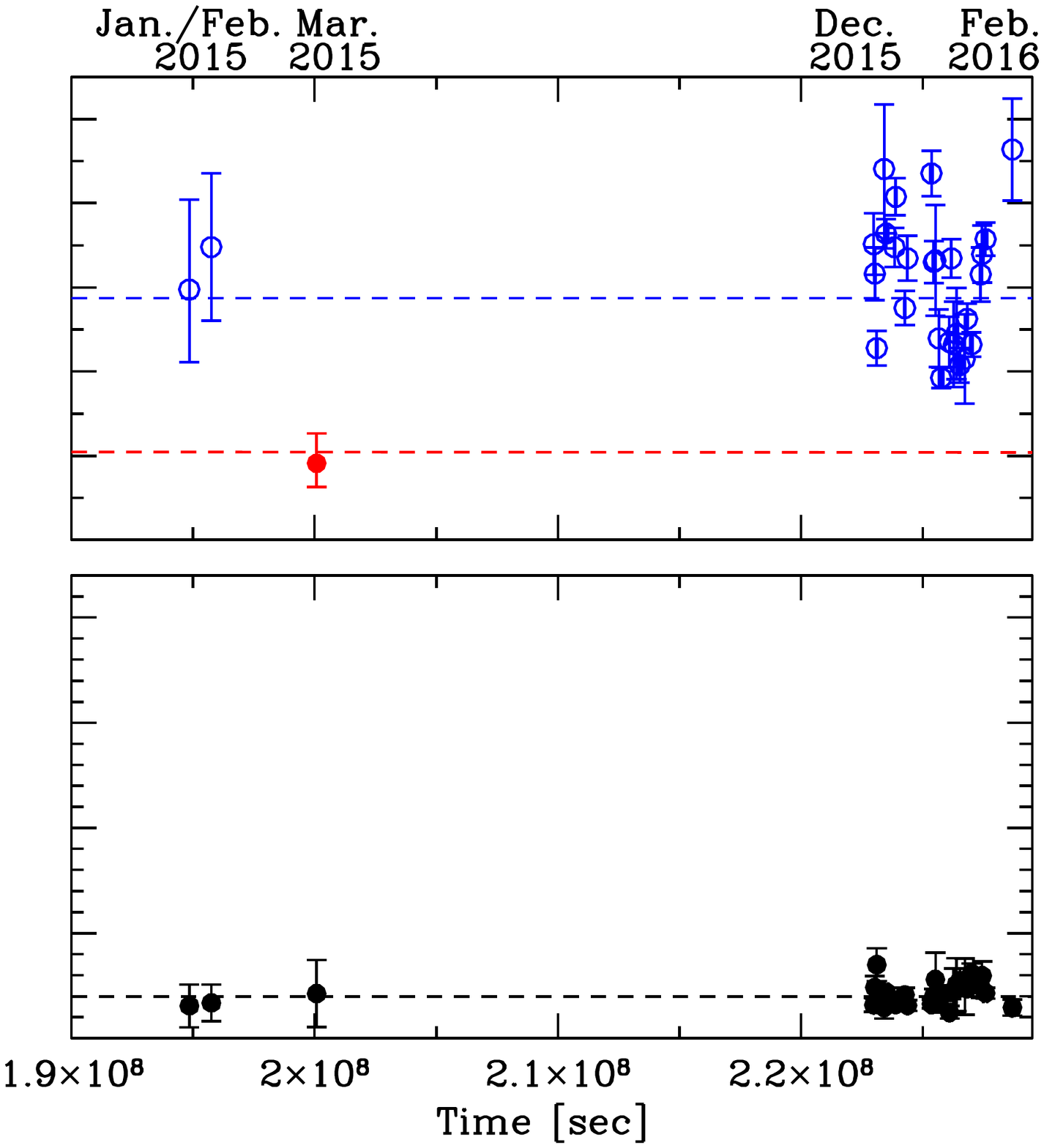}
    \caption{0.3-10 keV XRT count rate (upper panels) and hardness ratio (lower panels) light curves of MCG+11-11-032 obtained by binning the data per observation. Data points are corrected for technical issues (i.e. bad pixels/columns, field of view effects and source counts landing outside the extraction region) following the recipes discussed by  \citet{eva07, eva09}. Error bars mark 1$\sigma$ uncertainties, while dashed lines represent the weighted averages of the high and low count rate states (upper panels, blue and red lines, respectively) and of the hardness ratios (lower panels, black line). In the upper panels, blue and red points flag higher and lower count rate states, respectively.}
    \label{xrt_curve}
\end{figure*}
 \begin{figure}
 \begin{center}
    \includegraphics[scale=0.4,clip=true,trim=0cm 10cm 0cm 3cm]{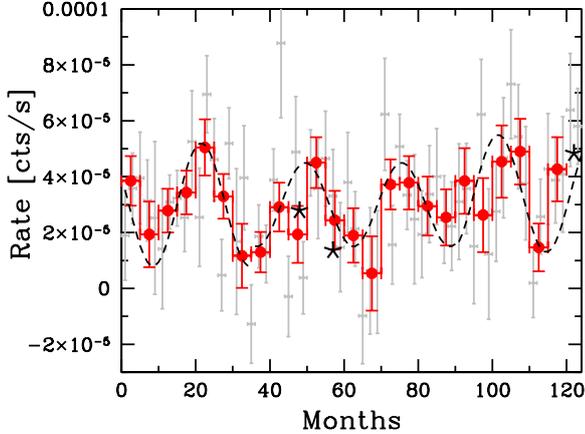}
    \caption{15-150 keV light curve of MCG+11-11-032  taken from the {\it Swift}-BAT 123 month survey (2005 January to 2015 March) in time bins of two (grey data points) and five months (red filled circles).  Error bars mark 1$\sigma$ uncertainties. The black skeletal symbols represent the binned XRT count rates overlapping in time with the BAT monitoring and rescaled to BAT count rates (see Sect. 3.2). For visual purposes only, we over-plotted a modular function obtained by summing four sinusoidal components with
equal period but different amplitudes (dashed black curve).}
    \label{bat_curve}
\end{center}
\end{figure}

\subsection{XRT light curve}
We first investigated the  source variability behavior  from year/month to day timescales.
We compared  the results obtained by our own monitoring program, probing variations on relatively short timescales (day-weeks),  
with  archival data, which allow us to explore  longer timescales (up to several years, see Table~\ref{xobs_log}).

Figure \ref{xrt_curve} shows the 0.3-10 keV XRT light curve (upper panels) and  the  hardness ratio light curve (bottom panels) of  MCG+11-11-032. 
This latter is defined as the ratio between the 4-10 keV  and the  2-4 keV count rates; as showed in \cite{bal15a}, this ratio
provides a strong indication of the amount of absorption.
Indeed, while  below 
$\sim$2 keV different soft components (soft excess, reflection, scattering, etc) can be present, in the 2--10 keV range, 
the spectrum of an AGN can be approximated, at the first order, by an absorbed power law.

The data are  binned per observation, and we  considered
all the observations listed in Table \ref{xobs_log}. Our daily monitoring data correspond to the data-points crammed in to the rightmost part of the
figure. The overall pattern of the count rate light curve
is not constant at the 99.99 per cent confidence level ($\chi^2$ test).
In particular, on month/year timescales the  source was caught  in two significantly different count rate
levels: the higher one (blue empty symbols, Fig.~\ref{xrt_curve}, upper panels), characterized by a 
weighted average count rate of $\sim$0.03 cts s$^{-1}$
(marked with a blue dashed line), and the lower one 
(red filled symbols, Fig.~\ref{xrt_curve}, upper panels), characterized by a  weighted average  count rate of $\sim$0.01 cts s$^{-1}$ (marked with a red dashed line).
The top-left panel of Fig.~\ref{xrt_curve} clearly shows that the source significantly increases its
count rate level in two months, and after about seven months the source has the
same lower  count rate value of the first observation. 
A similar but opposite trend was observed about seven years later (top right panel of Fig.~\ref{xrt_curve}):
the source decreases  its
count rate level in two months
 and, after about nine months, the source has already increased again  its count rate 
 to the higher state.
Unfortunately, the   lack of an  uniform sampling across the full period prevents us from further investigating this  ``alternating" behavior.
 The two states shown in Fig.~\ref{xrt_curve} are, most probably, 
 the states where the source spends the  majority of its time.

The overall pattern of variability observed in the full 0.3--10 keV  band, and shown in the upper panels of Fig.~\ref{xrt_curve}, is similar to that 
registered in the 2-4 keV and 4-10 keV energy ranges. 
Indeed, the hard-to-soft flux ratio,  plotted in the bottom panels of Fig.~\ref{xrt_curve}, does not show significant variations ($<$3$\sigma$, $\chi^2$ test),
except from the third observation in the left panel (Sep. 2009). 
The lack of significant variation in the count rate ratio suggests that the observed count rate variability  is 
not caused by variable absorption  but it is most likely 
due to intrinsic flux variations  (see also Sect. 3.3).

\subsection{BAT light curve}
To further investigate the month/year variability behavior shown by the XRT data,  
we considered the (still unreleased) MCG+11-11-032 123-month 15-150 keV BAT light curve 
\citep[see Fig.~\ref{bat_curve}; Palermo {\it Swift}-BAT team, private communication; see also][]{seg10},  
which provides a tight sampling of the source count rate on a total time scale of more than ten years.
In Fig.~\ref{bat_curve} we show the light curve binned in periods of two (grey data points) and five (red filled circles) months.
The 15-150 keV emission  is clearly variable;
a constant flux is rejected at 99.95 per cent confidence level ($\chi^2$ test).
The curve has a modular behavior with different peaks and dips occurring almost every 25 months.
For visual purposes only, we over-plotted on Fig.~\ref{bat_curve} a modular function obtained by summing four sinusoidal components with
equal period but different amplitudes (dashed black curve).

 In order to compare the XRT and BAT variability, we binned the XRT light curve in periods of five months
and then converted the XRT to BAT count rates{\footnote{We considered the XRT data obtained during the123-months of BAT
observation and we convert them to BAT count rates by using 
the PIMMS tool (v. 4.8f) and considering  the spectral parameters ($\Gamma$, N$_{\rm H}$) derived by our own spectral analysis reported in Sec. 3.3.}}
(skeletal symbols in Fig.~\ref{bat_curve}). We found a good  agreement between  the two datasets.

\subsection{XRT spectral analysis}

\begin{figure}
        \includegraphics[scale=0.33,clip=true,trim=0cm 0.8cm 0cm 2.5cm]{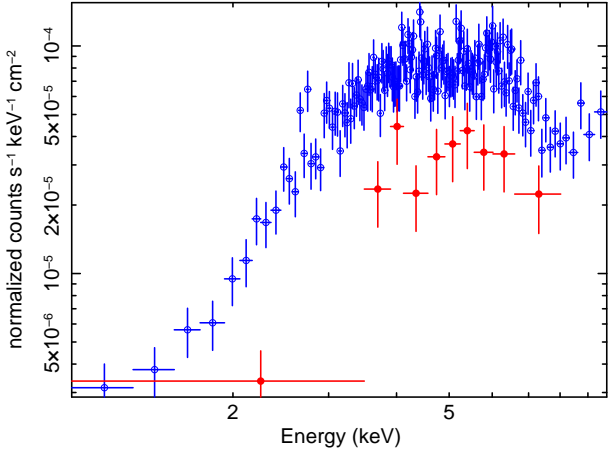}
    \caption{XRT spectra corresponding to the high (blue open circles) and low (red solid circles) states shown in upper panels
    of Fig.~\ref{xrt_curve}.}
    \label{ldata_area}
\end{figure}

\begin{table*}
 \begin{minipage}[t]{1\textwidth}
        \begin{center}
        \caption{Best-fitting values obtained by applying to the  high-state spectra (empty blue circles in Fig.~\ref{xrt_curve}) of MCG+11-11-032 the different models discussed in Sect. 3.3 (referenced 
        as models from 1 to 4). Errors are quoted  at the 90 per cent confidence level for one parameter of interest \citep{avn76}.}
            \label{xrays}
        \begin{tabular}{ccccccccc} 
                \hline
                \hline
                {\it Model} & $\Gamma$ & {\it N$_{\rm H}$} & {\it R} & {\it E} & {\it EW} & $\chi^2/d.o.f.$  & {\it F}$_{\rm {2-10 keV}}$ & {\it L}$_{\rm {2-10 keV}}$ \\
                                 &                    & [$10^{22}$ cm$^{-2}$] &  & [keV] & [eV] &                              & [10$^{-12}$ erg cm$^{-2}$ s$^{-1}$] & [10$^{43}$ erg s$^{-1}$] \\
                     (1)          & (2) &(3) & (4) & (5) & (6)  &(7) &(8) & (9) \\
                \hline
                1 &  1.29$^{+0.19}_{-0.18}$ & 11.0$^{+1.2}_{-1.5}$ &                                &                                      &                               & 133 / 75  & 4.39 & 2.15\\
                \hline
                2 & 1.61$^{+0.17}_{-0.18}$ & 13.5$^{+1.4}_{-1.5}$ &  $\sim$0.09              &                                       &                              & 95 / 74 & 4.32    & 2.09  \\
               \hline
                3 & 1.64$^{+0.19}_{-0.18}$ & 13.3$^{+1.5}_{-1.4}$ & $\sim$0.09 & 6.18$^{+0.09}_{-0.08}$ & 120$^{+50}_{-60}$& 83 / 72 & 4.27 & 2.05\\
               \hline
                \multirow{ 3}{*}{4} & \multirow{ 3}{*}{1.68$^{+0.19}_{-0.18}$} &\multirow{ 3}{*}{13.3$^{+1.5}_{-1.4}$} & \multirow{ 3}{*}{ $\sim$0.09} & 6.16$^{+0.08}_{-0.07}$ & 120$^{+80}_{-60}$& \multirow{ 3}{*}{79 / 70} & \multirow{ 3}{*}{4.24} & \multirow{ 3}{*}{2.05} \\
                \\
                & &  & &6.56$^{+0.16}_{-0.15}$ & 85$^{+70}_{-25}$ && & \\ 
                \hline
                 \hline
       \end{tabular}
        \end{center}
 {\bf Notes:} Col. (1): Model number as referenced in the text (see Sect. 3.3).
 Col. (2): Power-law and reflection component photon index.
 Col. (3): Intrinsic column density.
  Col. (4): Reflection fraction, defined as the ratio between the 2-10 keV flux of the reflected and direct continuum components.
  Col. (5): Rest-frame energy centroid of the Gaussian  line.
  Col. (6):  Emission line equivalent width.
  Col. (7): $\chi^2$ and number of degrees of freedom.
  Col. (8): Observed flux (de-absorbed by Galactic absorption) in the 2-10 keV energy band.
  Col. (9): Intrinsic (i.e. absorption-corrected) luminosity in the 2-10 keV energy band.
   \end{minipage}
\end{table*}

\begin{figure}
\includegraphics[scale=0.33,clip=true,trim=0cm 0.8cm 0cm 2.5cm]{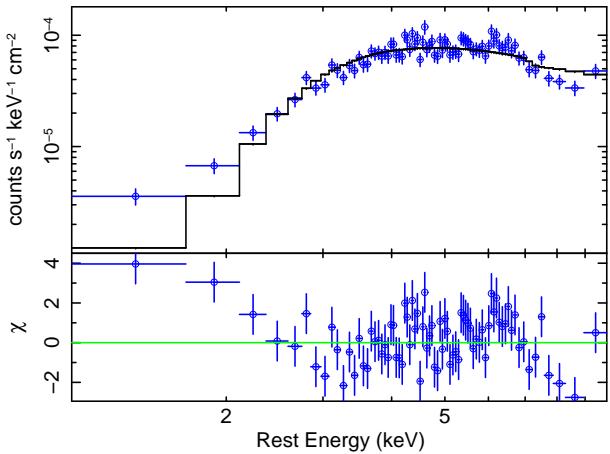}
    \caption{Upper panel: The model composed of an intrinsically absorbed power-law   ({\it tbabs*ztbabs*zpowerlw}, model 1, Table~\ref{xrays}) 
    is plotted over the XRT spectrum of MCG+11-11-032.
     Lower panel: Relevant residuals plotted in terms of sigmas.}
    \label{model1}
\end{figure}

\begin{figure}
 \includegraphics[scale=0.33,clip=true,trim=0cm 0.8cm 0cm 2.5cm]{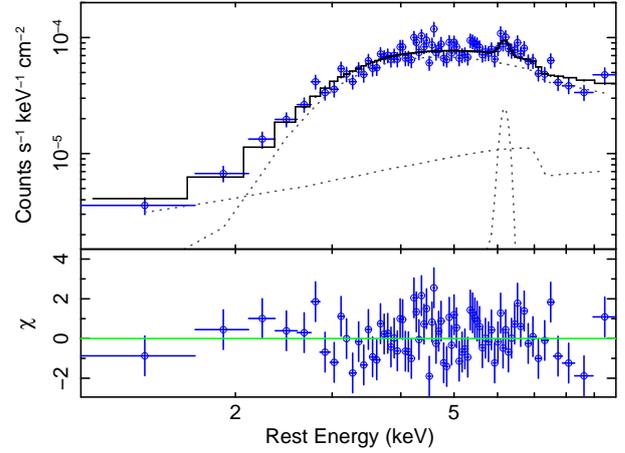}
    \caption{Upper panel: The model composed of an intrinsically absorbed power-law plus a continuum reflection component and one narrow emission  
    line ({\it tbabs*}({\it ztbabs*zpowerlw+pexrav+zgauss})),  model 3 in Table~\ref{xrays}) is plotted over the  XRT spectrum of MCG+11-11-032.
     Lower panel: Relevant residuals plotted in terms of sigmas.}
    \label{model3}
\end{figure}

\begin{figure}
        \includegraphics[scale=0.33,clip=true,trim=0cm 0.8cm 0cm 2.5cm]{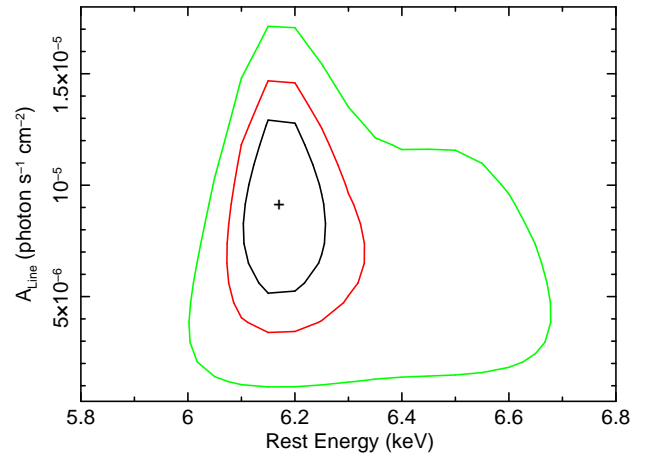}
            \caption{Confidence contours plot of the joint errors of the rest-frame emission line energy versus the 
 intrinsic normalization of the Gaussian component ({\it A$_{\rm line}$}). 68 per cent (black line), 90 per cent (red line) and 99 per cent (green line)
 confidence contours are shown.} 
 	\label{cont}
 \end{figure}
 
\begin{figure}
        \includegraphics[scale=0.33,clip=true,trim=0cm 0.8cm 0cm 2cm]{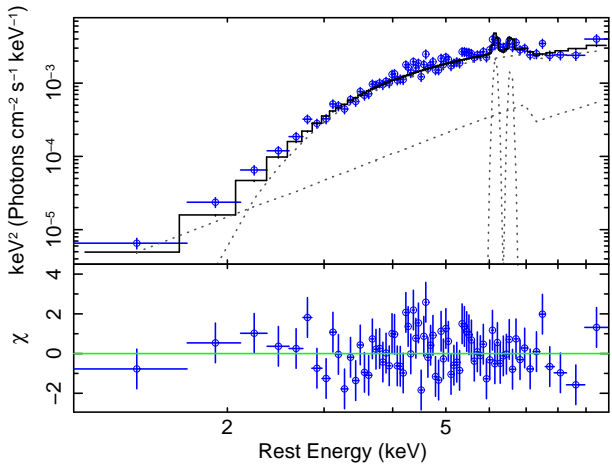}
    \caption{Upper panel: the model, which includes an intrinsically absorbed power-law plus a continuum reflection component and two emission narrow lines 
    ({\it tbabs*}({\it ztbabs*zpowerlw+pexrav+zgauss+zgauss})), model 4 in Table~\ref{xrays}), is plotted over the spectrum of MCG+11-11-032. 
    Note that this plot was obtained by creating fluxed spectrum against a simple $\Gamma$ = 2 power-law and then overlaying the best fit model.
    Lower panel: relevant residuals, plotted in terms of sigmas.}
    \label{model4}
\end{figure}

As already discussed in Sect. 3.1, by considering the XRT count rate ratios (see Fig.~\ref{xrt_curve}, lower panel), 
no  evidence of spectral variability is observed between the two states. 
To further test this result,
we produced two different spectra on the basis 
of the 0.3-10 keV light curve: we 
co-added all the data relevant to the blue  and red points showed in the
upper panels of Fig.~\ref{xrt_curve}  to obtain source and background spectra
of the high and low-flux states, respectively.
The resulting spectra are compared in Fig.~\ref{ldata_area}.
Due to the different statistical quality of the two spectra,
the high-state spectrum ($\sim$4170 net counts) has been binned in order to have at least 50 counts per energy channel,
while the low-state one  ($\sim$100 net counts) has been binned in order to have at least 10  counts per energy channel.
As evident from Fig.~\ref{ldata_area}, the two states have very similar spectral shapes as already suggested by the X-ray  colours.
Therefore, we can rule out variable absorption to be at the origin of the observed flux variability.
In the following, we will focus on the higher state and higher statistics spectrum.
The spectral analysis is performed by using the {\textsc{{\small XSPEC}}} 12.8.2
package \citep{arn96}. We use  the $\chi^2$  statistics in the search for the best fit model
and for parameter errors determination \citep{avn76};  quoted statistical
errors are at the 90 per cent confidence level for one parameter of interest.

Each model discussed below includes a Galactic column density {\it N}$_{\rm {H Gal}}$=4.7$\times$10$^{20}$ cm$^{-2}$
\citep{kal05}, modeled with {\it tbabs} in {\textsc{{\small XSPEC}}} \citep{wil00}.
Since the shape of the spectra in Fig.~\ref{ldata_area} suggests the presence of obscuration, as a starting point 
we adopted a single intrinsically absorbed power-law model ({\it ztbabs*zpow} model in {\textsc{{\small XSPEC}}},  model 1, Table~\ref{xrays}). 
As shown  in Fig.~\ref{model1}, such model  cannot be considered a good 
representation of the global spectral properties of the source leaving  evident
residuals over the full energy range. In particular, both the softer (below $\sim$ 2 keV) and the harder (above $\sim$6 keV) residuals suggest the 
presence of a non-negligible additional component tracing  reflection, most probably due to the 
circum-nuclear material.
We added to the fit first a continuum reflection component ({\it pexrav} model in  {\textsc{{\small XSPEC}}}, \cite{mag95},
 model 2, Table~\ref{xrays}) and then a narrow (50 eV) Gaussian emission line component around 6.4 keV to account for a Fe K$\alpha$  emission component ({\it zgauss}
 model in  {\textsc{{\small XSPEC}}},
 model 3, Table~\ref{xrays}). Both  components are statistically required and their addition  significantly improves the fit. 
The   spectrum and the residuals corresponding to the final  best-fitting model  (model 3, Table~\ref{xrays})
  are shown in Fig.~\ref{model3}, upper and lower  panels, respectively.
We note that the best-fit value of the rest-frame energy  (6.18 keV) of the narrow emission line does not have a clear association with well-known and expected transitions.
The difference with respect to the neutral  rest-frame Fe emission line at 6.4 keV is significant at 97 per cent confidence level
 for one parameter  of interest. 
 Fig.~\ref{cont} shows the confidence contour plot of the joint errors of the rest-frame emission line energy versus its 
 intrinsic normalization.  This figure indicates that a rest-frame energy of 6.18 keV is favored by the model, although also a 6.4 keV value, i.e.
the rest-frame Fe K$\alpha$ line, cannot be excluded at 3$\sigma$. The shape of the 99 per cent contour level hints also another
possibility: the presence of a second emission line at energy higher than 6.4 keV.
 For this reason,
 even if it is not required by the fit with the present statistics,
  we add a second narrow emission line to the model
 leaving its energy free to vary in 6-7 keV energy range (model 4, Table \ref{xrays}).
 The best fit values obtained for the rest-frame energies of the two emission lines are: {\it E$_{\rm 1}$}=6.16$\pm$0.08 keV ({\it EW}$\sim$120 eV, {\it F}$_{\rm line1}$$\sim$8$\times$10$^{-14}$ erg s$^{-1}$ cm$^{-2}$)  and {\it E$_{\rm 2}$}=6.56$\pm$0.15 keV ({\it EW}$\sim$85 eV,  {\it F}$_{\rm line2}$$\sim$6$\times$10$^{-14}$ erg s$^{-1}$ cm$^{-2}$).
As expected, due to the statistics of our data, the energy of the second line is poorly constrained.
  All the continuum parameters are unaffected with respect to those obtained by adopting model 3 (see Table~\ref{xrays}).
  For completeness, we report in Fig.~\ref{model4} the spectrum and the ratio between data and this last best-fitting model.

As a final step, we  check if the two putative X-ray emission lines could be associated with the double-horn of a relativistic 
Fe K$\alpha$ emission line produced in the accretion disk, i.e. inside the  absorbing medium intercepted along the line of sight  
({\it N$_{\rm H}$}=1.33$\times$10$^{23}$ cm$^{-2}$, see Table \ref{xrays}).
To this end, the two {\it zgauss} components are replaced  with  an absorbed {\it laor}  \citep{lao91} disk line plus a  {\it pexrav} component in {\rm XSPEC}.
These represent the iron emission line plus continuum components reflected by the accretion disk and absorbed by the outer medium along the line of sight.
We find that the   symmetric double peaked profile observed for MCG+11-11-032 (both in terms of $\Delta E$ and emission line fluxes)
can be reproduced by this model only by assuming that a significant fraction of the line flux comes from
a Keplerian disk at $\sim$200-400 {\it R}$_{\rm G}$ from the central engine.
In the case of a single ionizing source, such range  corresponds  to the outer part of an accretion disk or to the region where optical broad emission lines are typically produced, i.e. the so called Broad Line  Regions (BLR). As we will discuss in the next Section, the distance quoted above matches well
also the inner radius expected for a circumbinary accretion disk for the presence of two sub-parsec scale SMBHs.
All the possible different scenarios will be discussed in the next section.

\section{Summary and Discussion}
Our analysis of the SDSS--DR13 optical spectrum of MCG+11-11-032 (see Sect. 2) confirms the 
presence of double-peaked profiles in all the strongest nebular emission lines, particularly for the [OIII] lines.
Emission line components at the systemic velocity of the
host galaxy  and/or broad 
wings have not been detected.   The velocity offsets between 
the blue and red narrow peaks are similar for all the emission line components 
and  the separation line ratios ($\Delta \lambda_{\rm blue}$/$\Delta \lambda_{\rm red}$) are consistent with one. 
Although  these properties make MCG+11-11-032
a good dual AGN candidate \citep{wan09}, alternative and  equally valuable physical scenarios could
account for the optical double-peaked narrow emission line components observed in this source, 
such as faint outflows/jets or gas kinematics within a single NLR. 

Interestingly, besides being characterized by double-peaked narrow emission lines in the
optical spectrum, MCG+11-11-032 also belongs to the all-sky survey {\it Swift}-BAT catalogues and was
 monitored several times with {\it Swift}-XRT (see Section 3). 
 Our analysis of the  XRT light curve shows that MCG+11-11-032 clearly alternates between two main flux states on a  time scale of several months.
Unfortunately, the XRT light curve data are not uniform and they are inadequate to unveil a possible periodic behavior;
 they only suggest that, if a modular behavior is present, it must have a period equal or larger than about one year.
To further investigate the X-ray  variability pattern of MCG+11-11-032, 
 we considered  the still unreleased 123-month 15-150 keV BAT light curve,  which provides a tight sampling of the source count rate on a total time scale of more than ten years.
The light curve  is clearly variable and shows a modular behavior with different peaks and dips occurring almost every 25 months.
The XRT data analysis (count rate ratio and low and high state spectrum comparison) suggests that 
the observed  X-ray variability  is  most likely caused by intrinsic flux variations rather than to a change of  obscuration
along the line of sight.

Our spectral analysis of the higher state shows that the XRT spectrum of MCG+11-11-032 is well fitted by an absorbed power-law plus a reflection component. 
While the continuum is reminiscent of a typical Seyfert 2 galaxy, we did not detect any neutral Fe K$\alpha$ emission line at the expected 6.4 keV rest-frame energy.
Although this would not make MCG+11-11-032 an outlier, what makes it  more interesting  is the possible presence of two emission lines  at rest-frame energies of:  {\it E$_{\rm 1}$}=6.16$\pm$0.08 keV  and 
{\it E$_{\rm 2}$}=6.56$\pm$0.15 keV.
While the 6.56 keV emission can be  considered only a tentative detection (2$\sigma$ significance),
we note that the $\sim$6.2 keV line is detected at high significance (more than 3$\sigma$) and
 is not consistent with  the rest-frame Fe K$\alpha$ energy of 6.4 keV at the 97 per cent confidence level.
 
\subsection{A binary SMBH at the center of MCG+11-11-032}
Although the results presented here need to be confirmed by higher quality X-ray data,
the putative modular X-ray variability combined with  the possible presence of two Doppler-shifted iron emission lines 
opens a further interesting possibility for MCG+11-11-032: the presence of two sub-pc scale SMBHs in the core of the source.
As a matter of fact, modular variations of the intrinsic flux in AGN constitutes an almost unique observational evidence  
for the presence of a binary SMBH at sub-parsec scale \citep[][and references therein]{cha18}. Numerous hydrodynamical simulation show that 
the large amount of dense gas in the central region of galaxies  hosting sub-parsec scale binary systems
can form a circumbinary accretion disk with an inner radius lower than two times the
binary separation. The circumbinary disk accretion mass rate is  expected to be modulated by the orbital period of the two SMBHs, hence  naturally producing
modulated X-ray variations \citep{dor13,gol14,far14}. 
Alternatively, modular variability in the presence of a SMBH pair  could be caused by relativistic Doppler-boosts of the emission
produced in mini-disks bounded to individual SMBHs \citep{dor15}. In this case the emission of the brighter mini-disk will be periodically Doppler-boosted,
and the observed time-scale of the modulated X-ray emission would correspond, also in this case, to the orbital period of the two
SMBHs.
As for MCG+11-11-032, by considering a total SMBH mass of log{\it (M/M$_{\odot}$})=8.7$\pm$0.3 \citep[derived from the CO velocity dispersion, see][]{lam17}\footnote{The SMBH mass for MCG+11-11-032, ID~\#434 in the online-only extended Table 8 of \cite{lam17}, has been derived from the velocity dispersion of the CO line. 
See the relevant paper for more details.}, under the hypothesis of a
SMBH pair, the observed modular time-scale (i.e. about 25 months)  would  imply a sub-pc separation between the two SMBHs,
with an orbital velocity of a few per cent of the speed of  light  ($\Delta v$$\sim$0.06{\it c}).
In spite of the statistical significance of the X-ray results presented here, what makes the sub-pc SMBH pair a valuable hypothesis in the case of MCG+11-11-032 is
the  complete agreement between the  orbital velocity derived from the BAT light curve and  the velocity offset derived by  the rest--frame $\Delta E$ between the two X-ray line peaks
in the XRT spectra data (i.e. $\Delta v$ of order of 3-10 per cent the speed of light, with a best fit value of about 0.06{\it c}).  
Interestingly, as discussed in the previous section, the two X-ray emission lines may be either reproduced by  rotational effects of a 
Keplerian disk at distance well consistent with the putative circumbinary disk ($\sim$200-400  {\it R}$_{\rm G}$) or by two different gas structures bounded  to individual SMBHs.

At  these small separations, binary SMBHs may stall for a significant fraction of the Hubble time \citep[see e.g.][]{col14},  much higher
of the typical galaxy merger time-scale \citep[see e.g.][]{boy08,hop10}. In these systems optical narrow lines would come from  a much larger zone enveloping the binary system \citep{beg80} by tracing the velocity of the post-merger galaxy. Under the binary hypothesis, the double-peaked emission lines observed in MCG+11-11-032 could be thus explained by gas kinematics related to a NLR moving at the velocity of the almost edge-on host galaxy disk at sub-kpc scale.

\subsection{Alternative scenarios}
Alternative physical scenarios proposed on  the
basis of the optical data alone (see Section 2) would unlikely  explain both the optical and the X-ray properties of this source.
In particular, the presence of two SMBHs at larger scale (e.g. sub-kpc distance) would be at odd with the distance and velocity
derived by the X-ray data. Similarly, although the presence of a precessing outflow/jet 
 would  explain the presence of optical and X-ray double-peaked lines, this hypothesis remains unlikely on the basis of the expected precession period for a jet with a single SMBH
of mass of 10$^8$ M$_{\odot}$, i.e. 10$^{2.2}$-10$^{6.5}$ years \citep[see][]{lu05}. On the other hand, shorter periods are possible
for jet precession related to a SMBH binary system \citep[][and reference therein]{gra15}.
The optical double-peaked profiles could be caused by almost edge-on NLR disk gas kinematics ionized by a single SMBH, where the X-ray emission lines may be produced by the outer parts of a co-planar accretion disk.
In this case,  modular X-ray behavior could be justified by  the presence of a warped accretion disk which modulates
the intrinsic luminosity as it precesses \citep{gra15}. 
However, even under this hypothesis, the precession time-scale of a self-gravitating warped disk around a single SMBH of mass of order of 10$^8$ M$_\odot$ is much larger ($\sim$50 yr) than the putative modular time scale  observed for MCG+11-11-032 \citep[see][]{tre14}.

\section{Conclusions}
Although higher quality X--ray data are mandatory to confirm and  better characterize the observed X-ray emission lines
and to confirm the X-ray modular behavior, the results presented here make MCG+11-11-032 a promising binary SMBH candidate.
 Due to the stringent spatial resolution requirements,  confirming the presence of a sub--parsec binary SMBH  is significantly challenging both with  present,
i.e. the Hubble Space Telescope, and  future new generation telescopes, i.e. the European Extremely Large Telescope and
 the James Webb Space Telescope. However, these new upcoming facilities will 
significantly improve our statistics on the binary/dual SMBH population thus providing the data for a detailed study of this class of objects and of the 
multi-wavelength properties of their host galaxies.
In addition, future spatially--resolved spectral observations
could confirm the origin of the double peaked optical emission lines, and the location of their emitting regions with respect to the central nucleus hosting one or possibly two SMBHs.

Although the  interpretation proposed here   is admittedly in part   speculative, this paper clearly shows the high capability of X-ray data 
in unveiling SMBH pair candidates also in obscured sources.
In particular, the still on going {\it Swift}-BAT all--sky monitoring will
allow us to investigate the hard X-ray  light curve of  MCG+11-11-032
 on even longer time scales   and  to  definitively confirm its periodic-like  behavior.
Higher quality X-ray spectra are also necessary to better characterize the observed X-ray emission lines and thus 
confirm on more solid  ground the scenario proposed here. 
MCG+11-11-032 is an intriguing source as it could be   the first case in which  X-ray data unveil the presence of a sub-parsec binary SMBH on 
the basis of a double-peaked Fe K$\alpha$ emission line. We note that, such kind of profiles will be easily detected with  
the advent of  the  X-ray calorimeters such the  one developed for XARM and {\it Athena}.

\section*{Acknowledgements}
We  thank the anonymous referee for the useful and constructive comments which improved the quality of the paper.
CC acknowledges funding from the European Union's Horizon 2020 research and innovation programme under the Marie Sklodowska-Curie grant  No 664931.
PS  thanks P. Saracco, M. Dotti and A. Wolter for the useful discussions and helpful suggestions.
This work is based on data supplied by the UK {\it Swift} Science Data Centre at the University of Leicester.
We made use of the Palermo BAT Catalogue and database operated at INAF - IASF Palermo.
The SDSS is managed by the Astrophysical Research Consortium for the Participating Institutions of the SDSS Collaboration 
(see http://www.sdss.org/collaboration/citing-sdss/).



\nocite{}
\bibliographystyle{mnras}
\bibliography{severgnini_binary_mnras} 







\bsp	
\label{lastpage}
\end{document}